\documentclass[]{JFM-FLM_Au}

\lefttitle{H. Mohamed and J. Sesterhenn}
\righttitle{Journal of Fluid Mechanics}

 \title{Side-wall wetting and linear stability of falling films}

\author{Hammam Mohamed\aff{1} \and  J{\"o}rn Sesterhenn\aff{1}}

\affiliation{ \aff{1} Lehrstuhl f{\"u}r Technische Mechanik und Str{\"o}mungsmechanik,  Universit{\"a}t Bayreuth, 95440 Bayreuth, Germany}
\corresau{J{\"o}rn Sesterhenn, \email{Joern.Sesterhenn@uni-bayreuth.de}}

\begin{document}
	\maketitle
	
\begin{abstract}
	We investigate the influence of side-wall wetting on the linear stability of  falling liquid films confined in the spanwise direction.   A biglobal stability framework is developed, capturing inertia, viscosity, gravity, capillarity, and geometric confinement.   The base flow exhibits a curved meniscus and a streamwise velocity overshoot near the side walls.  Linear stability analysis based on the Navier--Stokes equations is performed in two limiting regimes. In \textit{confined} channels, where spanwise confinement stabilizes moderate-wavenumber perturbations via side-wall boundary layers, wetting weakens this stabilization; as the contact angle decreases, the neutral curves shift towards the unconfined one-dimensional limit, thus wetting acts as a relative destabilizing mechanism. In contrast, in \textit{weakly-confined} channels where side-wall boundary layers do not provide confinement-induced stabilization, wetting produces a net long-wave stabilization ($k \rightarrow 0$), significantly increasing the critical Reynolds number. This effect strengthens as the contact angle decreases, indicating a competition between destabilizing inertia and  stabilizing wetting-induced capillary forces. The predicted long-wave stabilization effect is compared quantitatively with available experimental measurements, showing consistent trends and comparable magnitudes within the accessible parameter range. Perturbation eigenmode structures show that, in confined channels, the relative destabilization is associated with near-wall vortical structures induced by the meniscus elevation and velocity overshoot, which reduce effective viscous damping. In contrast, in weakly-confined channels, stabilization is consistent with interface tensioning through strong anchoring of the perturbations at the side walls.
\end{abstract}


\section{Introduction}	
Gravity-driven liquid films are thin layers of fluid that flow down an inclined surface due to gravity. These flows have long attracted great interest as they are highly relevant both to fundamental science and a wide range of applications, such as film coating and heat exchangers. Since these flows are inherently unstable, understanding the mechanisms that govern their stability is crucial for predicting and controlling their behavior. 

The classical stability of gravity-driven liquid films has been extensively studied under idealized conditions where the film is unbounded in the spanwise direction. In such a configuration,  the hydrodynamic  instability is primarily governed by the interplay between inertia, viscosity and gravity, while surface tension plays a minimal role in the long-wavelength limit. Under these conditions, the critical Reynolds number becomes independent of  capillary effects. This classical framework has formed the foundation of our understanding of isothermal falling-film instability, and is particularly applicable where spanwise effects are negligible. 

Nevertheless, when the flow is confined in the spanwise direction, the stability behavior deviates significantly from the classical predictions. The interaction between inertia, viscous and capillary forces and spanwise confinement introduces new complexities that are not yet fully explored.    In our recent work \citep{mohamed2023effect}, we investigated the influence of spanwise confinement on the linear stability of gravity-driven liquid films using a combined experimental and theoretical framework based on a temporal biglobal stability analysis.  	We showed that perturbations at moderate frequencies experience heavy damping caused by spanwise confinement, leading to fragmentation of the instability domain. This distinctive stabilization was found to be a result of two actions: one is the presence of different types of stability modes that are suppressed differently by the spanwise confinement, while the second reason is due to a global bifurcation between the stability modes.  We identified two different stability mode types, the Kapitza hydrodynamic mode, usually known as \textit{H-mode}, which is dominant at weak spanwise confinement, and a new wall-confined stability mode, which we named \textit{W-mode} and which dominates at strong spanwise confinement. These results demonstrate how geometric confinement restructures the linear stability landscape of film flows, offering new insights into the interaction between instability mechanisms and boundary conditions.

A key simplifying assumption in our earlier model was to neglect  wetting effects at the side walls.  While this  assumption simplifies the stability model and facilitates the study of  geometric confinement, it overlooked a physically important phenomenon, namely side-wall wetting. It is  shown analytically and experimentally that the meniscus at the side walls alters the local interface shape and can result in a velocity overshoot in the vicinity of the side walls \citep{scholle2001exact,haas2011side}.   Such modifications to the flow become  particularly important in confined channels or in systems with strong capillarity, where the interaction between capillarity and spanwise confinement is expected to play a crucial role in shaping the flow dynamics. 

To date, the influence of wetting on the stability of falling films has been investigated exclusively through experimental studies, with no theoretical or numerical studies directly addressing this aspect. For instance, \cite{vlachogiannis2010effect} and \cite{georgantaki2011measurements} demonstrated that, unlike classical theory, the onset of the long-wave instability can be shifted by strong wetting  effects, even when the channel width is orders of magnitude larger than the film thickness.  Their experiments with aqueous solutions of glycerol or isopropanol had Kapitza numbers on the order of a few thousand, $\mathcal{O} (10^3)$. The latter study attributed this behavior to a transverse long-range capillary attenuation mechanism originating from the wetting effects at the side walls. In addition, \cite{pollak2011side} further confirmed that wetting introduces localized capillary effects near the side walls that stabilize the flow even when surface tension is weak.  This was demonstrated by obtaining the stability neutral curves at different spanwise locations, and observing a stabilizing shift in the stability curves near the side walls. They also showed that the streamwise velocity overshoot has a minor destabilizing effect compared to the dominant stabilizing effect induced by wetting.  It is important to note, however,  that these experimental studies were limited to channels much wider than the film thickness, where the stabilizing effect of the spanwise confinement is absent. Relatively narrower channels were not considered, where wetting could interact with the confinement-induced stabilization. 

Despite these experimental results, a comprehensive theoretical or numerical framework is still lacking to describe how wetting and confinement jointly shape film stability. In this work, we address this gap by extending our previous biglobal stability framework \citep{mohamed2023effect} to incorporate wetting effects at the side walls.  By explicitly resolving the modifications on the base state and perturbations introduced by the wetting, we aim to quantify how wetting alters the confinement-induced stabilization, and examine whether the velocity overshoot could lead to flow destabilization.  This approach enables us to systematically investigate the interaction between contact angle, surface tension, and geometric confinement, thereby offering a more complete and physically accurate insight into the stability characteristics of gravity-driven liquid films.

This work is organized as follows. In §\ref{sec:Theore_form}, we present the non-dimensional governing equations and  introduce the relevant dimensionless parameters. The derivation of the steady state solution and the linear stability problem are also included in the same section.  Section \ref{sec:results} presents our results and discussion. Finally,  section \ref{sec:conclusion} presents our concluding remarks and suggests potential directions for future investigations.  

\section{Theoretical formulation}\label{sec:Theore_form}
\begin{figure}
	\centerline{\includegraphics{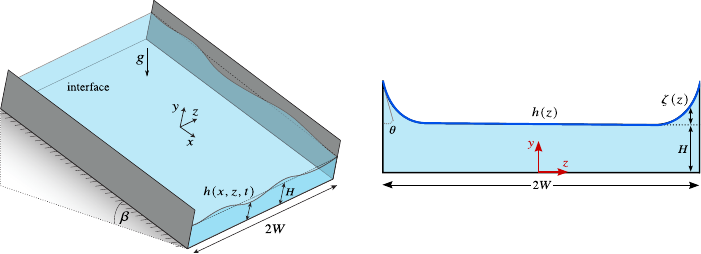}}
	\caption{Schematic diagram of a liquid film falling down an inclined channel. $h(x,z,t)$ is the local film thickness, $H$ is the mean film thickness, and $\zeta(z)$ is the capillary elevation due to wetting effects.}
	\label{fig:fig1}
\end{figure}
This section presents the theoretical foundation of this work, consisting of the governing equations, the appropriate non-dimensional scaling, the base state solution, and the linear stability problem. Figure  \ref{fig:fig1} shows a three-dimensional liquid film flowing down an inclined surface. The channel width is $2W$, where the side walls are located at $\pm W$, and inclined at an inclination angle $\beta$. A Cartesian coordinate system ($x,y,z$) is used, where $x$ denotes  the streamwise direction, $y$ is normal to the bottom wall, and $z$ represents the spanwise direction. More importantly, the flow is subject to wetting, with $\theta$ denoting the contact angle at the side walls. The film thickness far from the side walls is $H$, while  the capillary elevation due to wetting is given by $\zeta(z) $. Thus, the local film thickness is given by ${h}(z) = H + \zeta(z)$.   We now introduce the following scales in order to non-dimensionalize the governing equations and boundary conditions:
\begin{center}
	\begin{tabular}{ccc}
		$ \quad	h\rightarrow H \ h^*,\quad$ & $\quad(x,y,z) \rightarrow H \ (x^*,y^*,z^*),\qquad$ & $ \quad W\rightarrow H \ W^*,\quad$\\ \\
		$\quad t  \rightarrow H^2 \rho / \mu \ t^*,\quad$ & $\quad (u,v,w)  \rightarrow \mu / (\rho H) \ (u^*,v^*,w^*),\quad$   &  $ \quad p  \rightarrow  \mu^2 / (\rho H^2) \ p^*.\quad$\\ \\
	\end{tabular}
\end{center}
By using these scales and {dropping the asterisks for simplicity}, we obtain the dimensionless governing equations, namely, the continuity and Navier--Stokes equations, as follows:

\begin{subeqnarray}
	&   \partial_x u + \partial_y v + \partial_z w 	 =0, \\
	&   \partial_t u +\boldsymbol{u} \cdot \nabla {u}   	= - \partial_x p + \nabla^2 {u} + Re , \\
	&  \partial_t v +\boldsymbol{u} \cdot \nabla {v}  = - \partial_y p + \nabla^2 {v} - Ct , \\
	&  \partial_t w +\boldsymbol{u} \cdot \nabla {w} = - \partial_z p + \nabla^2 {w},
	\label{governing}
\end{subeqnarray}
where $\boldsymbol{u} = (u,v,w)$ is the velocity field, $p$ is pressure, $Re = g  \sin(\beta) H^3 /  \nu^2  $ is the Reynolds number based on the Nusselt film solution \citep{nusselt1916oberflachenkondensation}, and $Ct = Re \cot(\beta)  $ is the inclination number.

The boundary conditions at the free surface at $y=h(x,z,t)$ are described as follows: 
\begin{subequations}
	\begin{align}
		v &= \partial_t h + u  \partial_x  h + w \partial_z h,\\
		p &=  {2}\Big[   (\partial_x h)^2 \partial_x u + (\partial_z h)^2 \partial_z w + \partial_x h \partial_z h (\partial_z u + \partial_x w)  \nonumber  \\
		&\quad  -   \partial_x h (\partial_y u + \partial_x v)- \partial_z h (\partial_z v + \partial_y w) + \partial_y v\Big]/\hat{n}^2  \nonumber\\
		& \quad  - (Re/Ca ) \Big[ \partial_{xx} h(1+(\partial_z h)^2) + \partial_{zz}h (1+(\partial_x h)^2) - 2\partial_x h \partial_z h \partial_{xz} h \Big]/\hat{n}^3 ,\\
		0 &=      2\partial_x h (\partial_y v -  \partial_x u) + (1-(\partial_x h)^2)(\partial_y u + \partial_x v) - \partial_z h (\partial_z u + \partial_x w) \nonumber \\ 
		& \quad  -\partial_x h \partial_z h(\partial_z v + \partial_y w),\\
		0 &=      2\partial_z h (\partial_y v -  \partial_z w) + (1-(\partial_z h)^2)(\partial_y w + \partial_z v) -  \partial_x h (\partial_z u + \partial_x w) \nonumber \\
		&  \quad - \partial_x h \partial_z h(\partial_y u + \partial_x v) ,
	\end{align}
\end{subequations}\label{interfacecon}
where $ \hat{n} = [ 1+(\partial_x h)^2 + (\partial_z h)^2]^{1/2} $, and $Ca$ is the Capillary number which relates the viscous forces to capillary forces. Using the characteristic streamwise velocity $U_c = \rho g \sin (\beta) H^2/ \mu$ \citep{nusselt1916oberflachenkondensation}, the Capillary number can be defined as follows:

\begin{equation}
	Ca =  \frac{\mu U_c}{\sigma }=  \frac{ \rho g \sin(\beta) H^2 }{\sigma}.
\end{equation}
Although the Kapitza number $Ka$ does not appear explicitly in the governing equations, it is beneficial to introduce it, mainly to relate to experimental studies:

\begin{equation}\label{kapitza_eq}
	Ka = \frac{\sigma}{\rho  (g \sin\beta)^{1/3}  \nu^{4/3} } = \frac{Re^{2/3}}{Ca}.
\end{equation} 
where $Ka$ measures the relative importance of capillary to viscous-gravity force scales in the streamwise directions \citep{Chakraborty2014}. In the present configuration, we define the non-dimensional capillary length from the balance between the capillary and hydrostatic pressure in the direction normal to the plate, and use it to characterize the spanwise extent of the wetting effect at the side walls: 

\begin{equation}\label{capillary_length}
	l_c = \sqrt{\frac{\sigma}{\rho g \cos \beta}} / H.
\end{equation}

At the side walls,  the free surface meets the solid wall with an apparent contact angle $\theta$, which is imposed through the  geometric slope-angle relation: 

\begin{equation} \label{cl_bc}
	\frac{\partial h}{\partial z}\bigg|_{z=\pm W} = \pm \cot(\theta).
\end{equation}
In general, the contact angle $\theta$ is governed by local contact-line physics, such as dissipation, slip and hysteresis and may depend on the contact-line motion \citep[e.g.][]{hocking1987damping,viola2018capillary}. In the present work, $\theta$ is prescribed as the static contact angle $\theta_s$ in the base state, while the treatment of the contact angle perturbation in the linear problem depends on the adopted order of approximation and is specified in \S\ref{linear_Stability}.  

With respect to the wall boundary conditions, we apply a standard no-slip boundary condition at the bottom wall ($y=0$):

\begin{equation} \label{noslip_bc}
	u = v = w = 0.
\end{equation}

With regard to the side walls ($z=\pm W$),  the classical no-slip condition yields unbounded stress \citep{huh1971hydrodynamic,davis1974motion}, due to the well-documented singularity arising from a freely moving contact line over the solid substrate. This issue is commonly resolved by replacing the no-slip  condition by a slip boundary condition \citep{navier1823memoire,sui2014numerical}. Specifically, the tangential velocity at the side wall is assumed to be proportional to the local viscous stress at the wall, while a no-penetration condition is imposed in the normal direction:

\begin{equation} \label{slip_bc}
	u = l_s (\boldsymbol{\tau} \cdot \boldsymbol{n_z}) \cdot \boldsymbol{e_x} , \qquad v =  l_s (\boldsymbol{\tau} \cdot  \boldsymbol{n_z}) \cdot \boldsymbol{e_y}  , \qquad w = 0,
\end{equation}
where $l_s$ is the non-dimensional slip length regulating the slip condition, $\boldsymbol{\tau} $ is the non-dimensional viscous stress tensor, and $\boldsymbol{n_z}$ is the outward  normal vector at the side walls, defined as: 

\begin{equation} \label{non-stress}
	\boldsymbol{\tau} =  \nabla \boldsymbol{u} + \nabla^T \boldsymbol{u}, \qquad \boldsymbol{n_z}({\pm W}) = (0,0,\mp 1), 
\end{equation}
for an incompressible fluid and flat side walls, substituting equation (\ref{slip_bc}) into equation \eqref{non-stress} simplifies to at $z = \pm W$: 

\begin{equation} \label{slip_bc_final}
	u(\pm W)  =   \mp l_s \partial_zu  , \qquad v(\pm W) =   \mp l_s  \partial_z v , \qquad w(\pm W) = 0.
\end{equation}  

In summary, this formulation describes the essential physics of gravity-driven thin films confined in the spanwise direction and under side-wall-induced wetting. The governing equations and boundary conditions for the base state and the linear stability analysis are  presented in the following sections.

\subsection{Steady state solution}
The steady-state solution (base flow) represents the undisturbed flow state on which the linear stability analysis is based. In the present  configuration, the base state  is influenced not only by the spanwise confinement, but also by wetting effects at the side walls. This base flow has been studied analytically using asymptotic approaches \citep{scholle2001exact}, and experimentally by \cite{haas2011side}. Nevertheless, we compute the base state numerically, obtaining the interface shape and the associated velocity and pressure fields. 

We consider a two-dimensional base flow driven by the streamwise component of gravity along the plate. The flow is assumed to be  fully developed ($\partial_t=0$) and homogeneous in the streamwise direction ($\partial_x=0$). In the absence of any body force or pressure gradient in the spanwise direction, the base state z-momentum equation reduces to $\partial_z \bar{p} = 0$. Together with the homogeneity condition $\partial_x=0$, the continuity equation can be satisfied by $\bar{v} = \bar{w} = 0$. Thus, the variables associated with this base flow are taken as:

\begin{equation}
	u = \bar{u}(y,z), \quad \bar{v} = 0, \quad \bar{w}=0, \quad p = \bar{p}(y), \quad h = \bar{h}(z),
\end{equation}
which reflects the absence of any motion in the spanwise or bottom wall-normal direction. Under these assumptions, the governing equations  \eqref{governing} and boundary conditions \eqref{interfacecon} - \eqref{slip_bc_final} simplify to the following:

\begin{subeqnarray} \label{base_governing}
	& \partial_{yy} \bar{u} + \partial_{zz}\bar{u}  + Re = 0,    \\ 
	&\partial_y \bar{p} + Ct= 0, \\
	&\partial_z \bar{p} = 0,
\end{subeqnarray} 
with the interface boundary conditions at $y = \bar{h}(z)$: 

\begin{subeqnarray}\label{base_interface}
	& \partial_y \bar{u} - d_z \bar{h}  \partial_z \bar{u}= 0,   \\ 
	& \bar{p} + \frac{Re }{Ca} \Big[1 + (d_z \bar{h})^2\Big]^{-3/2}  d_{zz} \bar{h} = 0.
\end{subeqnarray}
At the side walls, the contact line  condition is: 

\begin{equation}
	\frac{d \bar{h}}{\partial z}\bigg|_{z=\pm W} = \pm \cot(\theta_s).
\end{equation}
The bottom  and side walls  velocity conditions are:

\begin{subeqnarray}
	& \bar{u}(y=0,z) = 0,   \\ 
	& \bar{u}(y,z=\pm W) =  \mp l_s \partial_z\bar{u}.
\end{subeqnarray}

The governing equations for the streamwise velocity (\ref{base_governing}a) and pressure (\ref{base_governing}b)  are coupled through the interface boundary conditions \eqref{base_interface} via the interface profile $\bar{h}(z)$, which motivates a more systematic approach to the problem. We proceed in two steps. First, we obtain the pressure and interface shape. Integrating equation (\ref{base_governing}b) and using the condition $\bar{p}|_{y={H}} = 0 $ gives: 

\begin{equation}
	\bar{p}(y) = Ct (H - y).
\end{equation}
Substituting this expression into the interface condition (\ref{base_interface}b) yields:

\begin{equation}  \label{middle_step}
	Ct(H - \bar{h}(z)) =-\frac{ Re}{Ca} \Big[1 + (d_z \bar{h})^2\Big]^{-3/2}  d_{zz} \bar{h}.
\end{equation}
We then write the interface profile as  $\bar{h}(z) = H +  \zeta(z)$. This leads to the governing equation for the capillary elevation $\zeta(z)$:

\begin{equation}
	\zeta  = \frac{Re}{Ca  Ct } \Bigg[1 + (d_z \zeta)^2\Bigg]^{-3/2} d_{zz} \zeta = \frac{Re}{Ca Ct} \  \frac{d}{dz} \Bigg[       \frac{d_z \zeta }{   \sqrt{  1 + (d_z \zeta)^2}  }    \Bigg],
\end{equation}
Multiplying by $ d_z \zeta$, and integrating by parts while using the conditions $\zeta(0) =  d_z \zeta(0) = 0$, we obtain:

\begin{equation}
	\zeta ^2 - \frac{6 Re}{Ca Ct}  \Bigg[ 1 -       \frac{1}{    \sqrt{ 1 + (d_z \zeta)^2}  }    \Bigg] = 0.
\end{equation}
After rearrangement, this can be written as the first-order problem:

\begin{equation} \label{interface_final}
	d_z \zeta = \Bigg[   \frac{1}{ (1 - \frac{Ca Ct}{6Re}  \zeta^2 )^2   } - 1     \Bigg]^{1/2}.
\end{equation} 
Equation (\ref{interface_final}) is then solved numerically to compute $\zeta(z)$ for a channel configuration in which the capillary elevation and its slope vanish at the center, consistent with the condition $\zeta(0) =  d_z \zeta(0) = 0$ used in the derivation. This corresponds to channels for which wetting introduces variations in the film thickness and velocity in the vicinity of the side walls, while the central part of the flow remains effectively one-dimensional. In practice, this  regime corresponds to $W/l_c \gtrsim 10$. For smaller ratios, the lateral menisci are no longer local and they overlap and reshape the film across the entire spanwise direction, requiring a different base state formulation. Within this range, equation~\eqref{interface_final} is solved numerically for $\zeta(z)$ using the boundary condition at
the walls:
\begin{figure}
	\centering
	\includegraphics[]{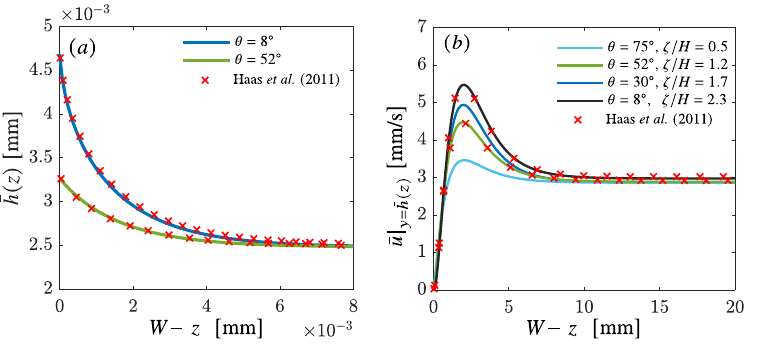}
	\caption{Base state solution: (a) Interface profile, and (b) Interface velocity in the vicinity of the side walls. Our results are compared 	with the experimental data of  \cite{haas2011side}}
	\label{interface_base}
\end{figure}

\begin{equation}
	\zeta(z=\pm W) = \Bigg[ \frac{2Re}{Ca Ct}  (1 - \sin(\theta_s)) \Bigg]^{1/2}.
\end{equation}
This leads to the interface profile $\bar{h}(z)$, which defines the physical domain of the base flow. Given $\bar{h}(z)$, the velocity and pressure fields are then obtained by solving the governing equations (\ref{base_governing}a,b) with the boundary conditions (\ref{base_interface}), and the wall conditions above. The resulting boundary-value problems for $\bar{u}(y,z)$ and $\bar{p}(y)$ are solved numerically using Chebyshev differentiation in a rectangular computational domain, to which the physical domain defined by $\bar{h}(z)$ is mapped. Further details of the mapping are provided in appendix \ref{Appendix_B}. 

Figure \ref{interface_base} presents the validation of our numerical solver  against the experimental data obtained by \cite{haas2011side}. Figure \ref{interface_base}(a) shows the interface profile near the side walls for different contact angle values and shows excellent agreement with experiments in both elevation amplitude and interface curvature. More importantly, the interface profile is accurately captured even for small contact angles ($\theta=8^{\circ}$), which corresponds to a higher capillary elevation. This notable increase in elevation is a result of the smaller radius of curvature, which increases the capillary forces pulling the liquid upward.

Moreover,  a distinctive feature of the base state solution is the velocity overshoot in the vicinity of the side walls. The capillary elevation induces a non-uniform film thickness in the spanwise direction. Under the same gravitational forces, the local Nusselt solution scales with the square of the film thickness, so the thicker regions support higher streamwise velocities. This spanwise variation of the film thickness leads to an overshoot in the streamwise velocity near the side walls \citep{zhou2020capillary}.  Excellent agreement is found between our  numerically  obtained velocity profile and the experiments of \cite{haas2011side}, as shown in figure \ref{interface_base}(b).  The magnitude of the velocity overshoot is proportional to the capillary elevation to base interface height ratio ($\zeta/H$). As shown, the velocity overshoot becomes more pronounced as this ratio increases, while the peak velocity shifts further away from the side walls. This emphasizes the important role of wetting effects and capillary elevation in shaping the velocity field near the side walls. 

\subsection{Linear stability analysis}\label{linear_Stability}
We perform a modal linear stability analysis based on the biglobal approach, which accounts for the variations in both the wall-normal ($y$) and spanwise directions ($z$), enabling the examination of the stability modes affected by spanwise confinement, and more importantly, wetting effects. The flow field is decomposed into a combination of the steady state and infinitesimal perturbations as follows: 

\begin{subeqnarray}
	&\boldsymbol{u}(x,y,z,t) = \bar{\boldsymbol{u}}(y,z) + \epsilon \tilde{\boldsymbol{u}}(x,y,z,t),\\
	&{p}(x,y,z,t) = \bar{{p}}(y) + \epsilon \tilde{{p}}(x,y,z,t),\\
	&	h(x,z,t) = \bar{h}(z) + \epsilon \tilde{h}(x,z,t),
\end{subeqnarray}  \label{expansions}
where  $\epsilon \ll 1$ is the perturbation amplitude. Substituting this expansion into the governing equations (\ref{governing}) and boundary conditions  (\ref{interfacecon})-(\ref{slip_bc_final}) and retaining the terms up to $O(\epsilon)$ leads to the linearized equations governing the perturbations, which are expressed as follows:

\begin{subeqnarray}\label{governing_per}
	& \partial_x \tilde{u} + \partial_y \tilde{v} + \partial_z \tilde{w} = 0, \\
	& \partial_t \tilde{u} + \bar{u} \partial_x \tilde{u} + (\partial_y \bar{u} )\tilde{v} + (\partial_z \bar{u}) \tilde{w} + \partial_x \tilde{p} - \nabla^2 \tilde{u} = 0, \\
	& \partial_t \tilde{v} + \bar{u} \partial_x \tilde{v} + \partial_y \tilde{p} - \nabla^2 \tilde{v} = 0,\\
	& \partial_t \tilde{w} + \bar{u} \partial_x \tilde{w} + \partial_z \tilde{p} - \nabla^2 \tilde{w} = 0.
\end{subeqnarray}
The linearized interface boundary conditions are as follows:

\begin{subequations}
	\begin{align}
		& \partial_t \tilde{h} + \bar{u} \partial_x \tilde{h} + \partial_z \bar{h} \tilde{w} - \tilde{v}= 0,\\
		&\tilde{p} + \partial_y \bar{p} \tilde{h} - {2} \big[  ( \partial_z \bar{h})^2  \partial_z \tilde{w} -  \partial_z \bar{h} (\partial_z \tilde{v} + \partial_y \tilde{w})   +    \partial_y \tilde{v}   \big] / \bar{n}^2  \nonumber  \\ 
		& \qquad \qquad \qquad \qquad  \qquad \qquad +  \mbox{}    (Re/Ca)   \big[   [1+ ( \partial_z \bar{h})^2]  \partial_{xx} \tilde{h} + \partial_{zz} \tilde{h} \big] / {\bar{n}^3} = 0, \\
		&   \partial_{yy} \bar{u} \tilde{h} + \partial_y \tilde{u} + \partial_x \tilde{v} - \partial_z \bar{u}  \partial_z \tilde{h} - \partial_z \bar{h} ( \partial_z \tilde{u} + \partial_{zy} \bar{u} \tilde{h} + \partial_x  \tilde{w} ) = 0, \\
		&  (1 - (\partial_z \bar{h})^2)   (\partial_y \tilde{w} + \partial_z \tilde{v}) - \partial_z \bar{u} \partial_x \tilde{h} - \partial_z \bar{h} \partial_y \bar{u} \partial_x \tilde{h} + 2\partial_z \bar{h} (\partial_y \tilde{v} - \partial_z \tilde{w})=0.
	\end{align}
\end{subequations}\label{interfacecon_linearized}
Notably, the interface boundary conditions are significantly more complex compared to those in the non-wetting case, as new terms proportional to $\partial_z \bar{h}$ appear. With regard to the contact line condition, contact line dynamics enter only beyond leading order linear stability analysis \citep{viola2018theoretical,snoeijer2007relaxation,savva2017asymptotic}. Accordingly, we impose no perturbation of the contact angle ($\tilde{\theta}=0$). Linearization of the geometric slope-angle relation \eqref{cl_bc} thus yields the free-edge boundary condition:

\begin{equation} \label{linear_cl}
	\frac{\partial \tilde{h}}{\partial z}\bigg|_{z=\pm W} =0.
\end{equation}
This free-edge condition is  the most suitable  closure for the present configuration. An alternative would be a pinned contact line, $\tilde{h} = 0$, which enforces zero displacement at the side walls and is typically associated with geometric pinning or brim-full containers \citep{kidambi2007oscillations}. In contrast, the present problem involves a partially filled channel with side walls with a prescribed static contact angle, for which the free-edge condition provides a more appropriate leading-order physical interpretation. 

The linearized boundary conditions at the bottom walls are: 

\begin{equation}\label{bottom_wall_per}
	\tilde{u} = \tilde{v} = \tilde{w} = 0,
\end{equation} 
while at the side walls ($z = \pm W)$, we have: 

\begin{subeqnarray}\label{side_wall_per}
	& \tilde{u} \pm l_s  \partial_z \tilde{u}= 0       ,\\
	&  \tilde{v} \pm l_s\partial_z \tilde{v}  = 0. \\
	& \tilde{w} = 0,
\end{subeqnarray} 

Additionally, we introduce the compatibility condition at the side walls for the pressure perturbation boundary condition \citep{theofilis2004viscous}:   

\begin{subeqnarray}\label{compatibility}
	&  \partial_y \tilde{p}  = \nabla^2 \tilde{v} - \bar{u} \partial_x \tilde{v},\\
	&  \partial_z \tilde{p}  = \nabla^2 \tilde{w} - \bar{u} \partial_x \tilde{w}.
\end{subeqnarray}
Finally, the perturbations are assumed to have the following form:

\begin{subeqnarray}\label{exp} 
	&  	\tilde{\boldsymbol{u}}(x,y,z,t) = \boldsymbol{\hat{u}}(z,y) \exp(ikx - i\omega t),\\
	&  	\tilde{{p}}(x,y,z,t) = {\hat{p}}(z,y) \exp(ikx - i\omega t),\\
	& \tilde{h}(x,z,t) = \hat{h}(z)  \exp(ikx - i\omega t).
\end{subeqnarray}  

We follow a temporal stability analysis where $k$ represents the wavenumber of the perturbations in the streamwise direction, $\omega$ is the complex angular frequency, where the imaginary part $\omega_i$ determines the temporal growth rate: $\omega_i > 0$ the perturbations grow and the flow is unstable and vice versa. 

\subsection{Numerical method}
Substituting the expansions (\ref{exp}) into the linearized governing equations and boundary conditions (\ref{governing_per}-\ref{compatibility}) results in a general eigenvalue problem. This problem is discretized using spectral collocation methods with Chebyshev polynomials in both the wall-normal ($y$) and spanwise ($z$) directions \citep{trefethen2000spectral}.    The boundary conditions are enforced by appropriately modifying the corresponding rows of the spectral matrices.

A two-dimensional mapping is implemented to transform  the non-rectangular physical domain into the rectangular computational Chebyshev domain, which accounts for the curved free surface associated with the meniscus. Further details on the numerical mapping are provided in appendix \ref{Appendix_B}. The resulting eigenvalue problem is solved using the QZ algorithm, where the solution consists of the eigenvalues ($\omega$), and the associated eigenfunctions ($ \boldsymbol{\hat{u}},  \hat{p}, \hat{h}$) which represent the perturbation amplitudes. 

With regard to the slip length $l_s$, prescribing a physically realistic molecular value is not computationally feasible, as molecular slip lengths are typically on the  nanometer scale \citep{eggers2004characteristic}, which is far below feasible numerical resolutions. Consequently, the slip length in the present formulation does not represent its exact molecular value, instead, it serves as an effective parameter that regularizes the singularity at the side walls, with its value linked to grid resolution \citep{renardy2001numerical,afkhami2009mesh}.  For the results reported here, the non-dimensional slip length is chosen to be $10^{-2}$, while the minimum grid spacing is fixed to $\Delta z_{min} = 1 \times 10^{-3}$.  Grid convergence tests and a detailed discussion on the sensitivity of the results to the slip length are presented in Appendix \ref{slip_length}. 

\section{Results}\label{sec:results}
The linear stability of gravity-driven liquid films is classically examined in a two-dimensional configuration where the spanwise direction is considered to be infinite, and the main controlling parameters are inertia, viscosity, gravity and, at finite wavenumber, capillarity.  When spanwise confinement is introduced, our previous study \citep{mohamed2023effect} has shown that the ratio of  channel half-width to mean film thickness ($W/H$) significantly modifies the stability behavior when $W/H \lesssim 50$, with notable stabilization of perturbations at moderate wavenumbers. This effect arises primarily from the development of a viscous boundary layer near the side walls,  which locally dampens perturbations and suppresses the hydrodynamic instability.  

In the present work, we investigate  the influence of side-wall wetting on the linear stability and its interaction with the stabilizing effect of the side-wall viscous boundary layers caused by spanwise confinement. We first consider a geometric confinement ratio of $W/H =20$, for which the viscous boundary layer thickness at the side walls is non-negligible compared to the film width, leading to the stabilizing effect described above. We therefore refer to $W/H = 20$ as a “confined” channel in the sense that side-wall viscous boundary layers are dynamically important. In contrast,  at $W/H = 100$, the instability of the flow is weakly influenced by the side-wall viscous boundary layers, and the stability characteristics are similar to those of a two-dimensional unconfined falling film limit. On this basis we refer to $W/H = 100$ as a “weakly-confined” channel, in which viscous effects due to spanwise confinement are negligible. 

In order to  quantify the spatial extent of the wall-induced wetting, we utilize the confinement to capillary length ratio $W/l_c$. We focus on channel configurations in which wetting introduces local modifications of the film thickness, curvature and velocity in the vicinity of the side walls, while the bulk of the flow behaves mainly as an unconfined falling film. For this configuration, the smallest ratio considered is $W/l_c = 10$, where smaller ratios lead to overlapping lateral menisci that reshape the flow across the entire width, defining a different physical regime that lies outside the scope of this work. A third key parameter is the contact angle $\theta$ at the side walls, which together with $l_c$, controls the meniscus shape,  and thereby influences  the stability characteristics. 

For clarification, the Capillary number is chosen as the primary control parameter for wetting effects. This method ensures a constant confinement-to-capillary-length ratio ($W/l_c$), which preserves the shape and influence of the wall induced meniscus across different flow regimes. The Kapitza number is also reported according to equation \eqref{kapitza_eq}, to offer qualitative comparison with experimental data which is typically based on $Ka$. This approach ensures both physical consistency in the modeling and relevance to experimental observations.

\begin{figure}[t]
	\centering
	\includegraphics{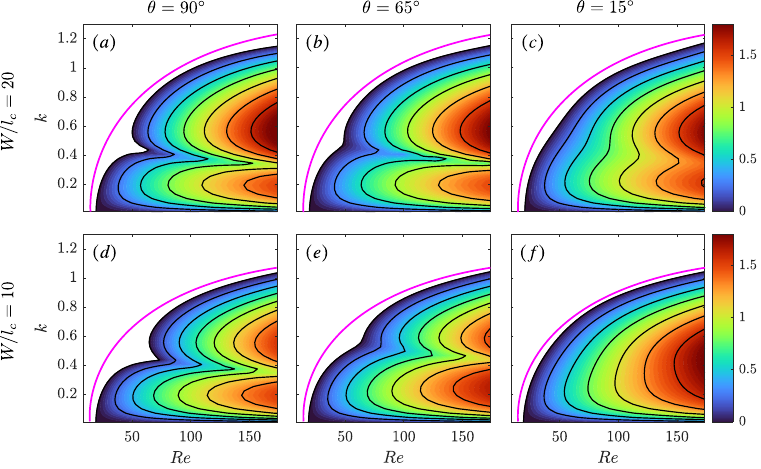}
	\caption{ 	Temporal growth rate contours in \( Re\text{--}k \) space for different values of $W/l_c$ and \( \theta \), with \( W/H = 20 \) and \( \beta = 10^\circ \). Panels (a–c) correspond to a Capillary number \( Ca = 0.15 \), while panels (d–f) correspond to \( Ca = 0.04\). The purple line represents the neutral  curve for the one-dimensional (unconfined) flow, corresponding to \( W \to \infty \). }
	\label{fig:narrowcontours}
\end{figure}
\subsection{Confined channels ($W/H = 20$)}
We begin our analysis by considering a  confined channel characterized by a confinement ratio of $W/H=20$. In this regime, the side walls introduce a strong viscous  boundary layer near the side walls that leads to confinement-induced stabilization. The  objective here is to investigate how wetting effects interact  with and potentially modulate this  stabilizing mechanism.  

Figure \ref{fig:narrowcontours} shows the temporal growth rate contours in the $Re-k$ plane for different values of the channel width to capillary length ratio  $W/l_c$  and contact angle $\theta$. The ratio $W/l_c$ is varied by adjusting the Capillary number $Ca$, while keeping the channel width $W$ fixed. \textcolor{black}{In terms of the Kapitza number, the present parameters correspond to $Ka = 100-300$, which lies within the range reported in earlier experiments \citep{georgantaki2011measurements}}. In the absence of wetting ($\theta=90^\circ$), the well-documented confinement-induced stabilization is observed for both $W/l_c$ values where perturbations at intermediate  wavenumbers are suppressed. For larger $W/l_c$ value corresponding to weaker capillary forces, the instability domain extends to higher wavenumbers, while the onset of the long-wave instability in the limit $k \rightarrow 0$ remains unchanged, consistent with classical theory in which surface tension plays no role in this limit. This is evident when comparing comparing panels (a) and (d).   With regard to the ridge that appears at moderate wavenumbers, it marks a change in the identity of the most unstable eigenmode. In this region of the $Re-k$ plane, different types of  instability modes have comparable growth rates, and since figure  \ref{fig:narrowcontours} shows only the leading eigenvalue, the switch from one branch to the other appears as a ridge rather than a smooth variation of the growth rate. This mode competition was examined in more detail and  discussed extensively in our previous work \citep{mohamed2023effect}.

\begin{figure}
	\centering
	\includegraphics{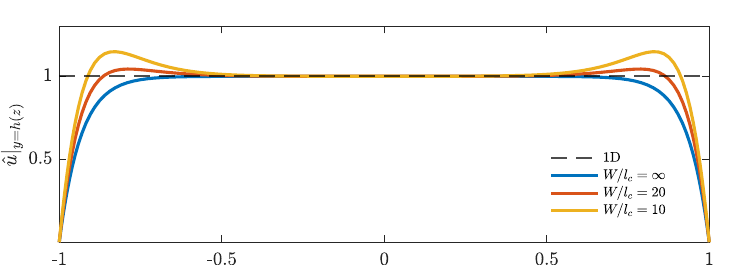}
	\caption{Normalized base state  streamwise velocity at $y = \bar{h}(z)$ for different $W/l_c$ values for $W/H=20$. }
	\label{fig:interfacevel}
\end{figure}

When wetting is introduced ($\theta = 65^{\circ}$), the stabilization effect of confinement at moderate wavenumbers is unexpectedly weakened.  The instability contours start to resemble those of the unconfined one-dimensional case, especially when surface tension is stronger (i.e., $W/l_c=10$). As the contact angle decreases further to $\theta=15^{\circ}$, the confinement-induced stabilization  is almost entirely eliminated as shown in panel (f). This behavior indicates that increasing wall wettability enhances capillary elevation near the side walls, which counteracts the stabilizing influence of the viscous boundary layer.    Importantly, while changes in $\theta$ do not alter the ratio $W/l_c$, they do modify the interface curvature and thereby the magnitude of the lateral capillary forces.

To better understand the influence of wetting on the confinement-induced stabilization, we examine the streamwise base flow velocity  for different $W/l_c$ values.  Figure \ref{fig:interfacevel} presents  the normalized base state streamwise velocity at the interface $\hat{u}= \bar{u}(y=h(z),z)/\bar{U}(H)$, where $\bar{U}(H)$ is the interface streamwise velocity in an infinitely wide channel.  When wetting effects are negligible ($ W \gg l_c $), the velocity profile exhibits a clear deceleration in the vicinity of the side walls, indicative of a viscous boundary layer which leads to flow stabilization. However, as wetting is introduced ($W/l_c$ decreases), the flow is accelerated near the side walls, causing a velocity overshoot which becomes more pronounced with smaller $W/l_c$ values, resulting in the reduction of the thickness of the stabilizing viscous boundary layer.

These findings suggest that, in confined channels, the presence of side-wall wetting can substantially reduce, and in some cases almost cancel, the viscous boundary layer induced stabilization provided by spanwise confinement. In this sense, side-wall wetting  plays a previously unreported relatively destabilizing role. Next, we turn to the weakly-confined channel case, in which the stabilizing effect of the side-wall boundary layers is negligible.

\subsection{Weakly-confined channels ($W/H = 100)$}
Next, we consider the case of weakly-confined channels, in which the spanwise confinement ratio $W/H$ is sufficiently large that the thickness of the side-wall boundary layer is negligible compared to the channel width and therefore, no confinement-induced stabilizing effect is present.   In this configuration, the stability behavior is effectively similar to that of the one-dimensional unconfined case, which allows us to isolate the role of wetting without the interference  from spanwise confinement.
\begin{figure}
	\centering
	\includegraphics{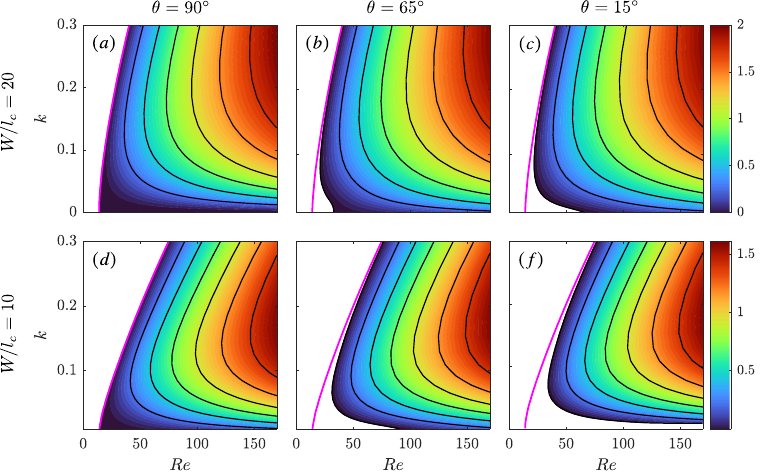}
	\caption{Temporal growth rate contours in the \( Re\text{--}k \) space for different values of $W/l_c$ and \( \theta \), with \( W/H = 100 \) and \( \beta = 10^\circ \). Panels (a–c) correspond to a fixed Capillary number \( Ca = 0.008\), while panels (d–f) correspond to \( Ca = 0.002 \). The purple line represents the neutral  curve for the one-dimensional (unconfined) flow, corresponding to \( W \to \infty \).}
	\label{fig:contours}
\end{figure}

Similarly, figure \ref{fig:contours} presents the temporal growth rate contours in the \( Re\text{--}k \)  plane for various values of $W/l_c$ ratio and contact angle.  For $\theta = 90^{\circ}$,  the unstable region narrows only at high wavenumbers as $W/l_c$ decreases, which is caused by the increase in surface tension. The onset of the long-wave instability remains unchanged, consistent with the classical one-dimensional theory. Furthermore,  the instability contours match those of the $1D$ unconfined case, confirming the absence of confinement-induced stabilization. 

As the contact angle decreases to $\theta = 65^{\circ}$, more pronounced wetting results in a contraction of the instability region at  small wavenumbers, leading to the suppression of long-wave perturbations. This leads to a shift in the onset of the instability to higher Reynolds numbers, while the growth rate at larger wavenumbers ($k\gtrsim 0.1$) remains largely unchanged.  This effect is more pronounced with a decreasing $W/l_c$ ratio.  A further reduction in contact angle to $30^{\circ}$  strengthens this effect, with long-wave instability being completely suppressed in the given parameter range under strong wetting conditions. The parameters presented here correspond to a Kapitza number in the range $Ka = 500 - 1500$, which lies within the values reported in earlier experiments \cite{georgantaki2011measurements}.

\begin{figure}
	\centering
	\includegraphics{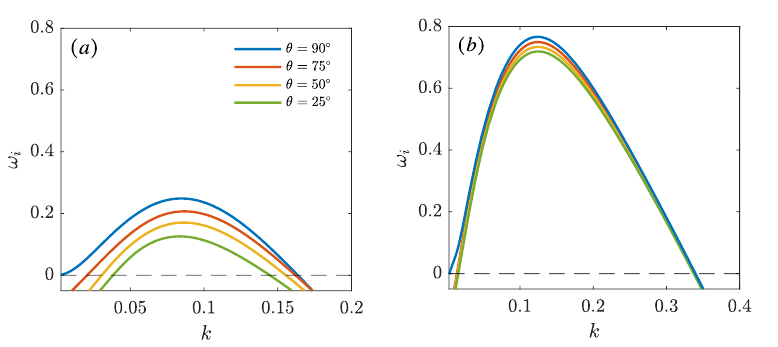}
	\caption{Temporal growth rate for different contact angle values for (a) $Re = 50$ and (b) $Re = 100$, when $W/H = 100$, $Ca = 0.01$, $\beta = 10$.}
	\label{omega_k_for_G}
\end{figure}

Additionally, figure \ref{omega_k_for_G} illustrates the temporal growth rate $\omega_i$ as a function of wavenumber $k$ at  two different Reynolds numbers. A clear stabilization is observed across all wavenumbers, particularly in the long-wave limit ($k \rightarrow 0$) at low Reynolds numbers $Re = 50$ shown in panel (a). This behavior is consistent with the wetting-induced stabilization observed in the contour plots. However, at higher Reynolds number ($Re=100$), shown in panel (b), the stabilization influence of wetting becomes significantly weaker at all wavenumbers. The growth rate curves for different contact angles nearly converge across all wavenumbers, indicating diminishing wetting stabilization.

In both cases, the wetting influence on the flow is similar since the Capillary number is fixed. The observed reduction in the stabilization is therefore not a result of weakening the wetting itself, but rather due to a shift of the dominant mechanism. 
As $Re$ increases, inertial effects become more dominant over wetting and the flow transitions into an  inertia-dominated regime that is less sensitive to wetting effects. This indicates that wetting is more effective at low $Re$ numbers, where capillary forces play a more dominant role over inertial forces. 

Taken together, the results highlight a distinct difference in the influence of wetting on the stability between confined and weakly-confined, despite maintaining a fixed $W/l_c$ ratio. In weakly-confined channels, stronger surface tension and larger capillary elevations are possible since the geometric restrictions on meniscus formation are weak. This provides room for strong wetting effects that lead to the stabilization of long-wave perturbations, shifting the instability onset to higher Reynolds numbers. 

On the other hand, in confined channels, the geometric constraint limits stronger wetting effects and prevents the strong stabilizing effect observed in wide channels. Instead, the velocity overshoot that develops near the side walls due to capillary elevation weakens the viscous boundary layer that drives the confinement-induced stabilization. Therefore, wetting acts as a relatively destabilizing effect with respect to the non-wetting confined case. In weakly-confined channels, however, the velocity overshoot is restricted to a small region near the side walls and thus has a negligible effect on the stability.  This interpretation is consistent with the experimental observations of \citet{pollak2011side}, who found that in wide channels, the destabilizing role of velocity overshoot is minor relative to the dominant stabilizing influence of surface tension.
\subsection{Transition between confined and weakly-confined channels}

\begin{figure} 
	\centering
	\includegraphics{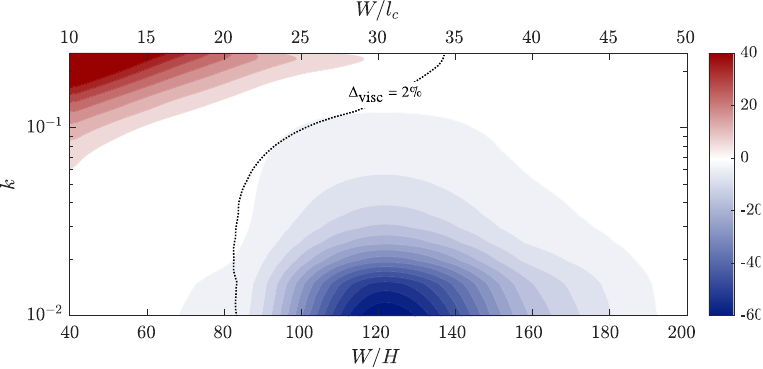}
	\caption{Phase diagram of wetting-induced modifications to linear stability neutral curve as a function of the wavenumber $k$ and  the confinement ratio $W/H$. Color contours show the percentage shift of the critical Reynolds number due to wetting according to \eqref{wet_shift}. Parameters: $\beta=5$, $Ca=0.056$, and  $\theta=15$. }
	\label{phase}
\end{figure}
The previous sections examined the effect of side-wall wetting in two limiting configurations. In a  confined channel ($W/H=20$), where viscous boundary layers at the side walls are dynamically important, wetting was found to have a relatively destabilizing effect on the flow, as it reduces the effectiveness of the viscous boundary layers that stabilize the flow at moderate wavenumbers. In contrast, wetting was found to have  little influence on the long-wave instability in the limit $k \rightarrow 0$.  For  a weakly confined channel ($W/H=100$), where viscous confinement effects are negligible, and the stability characteristics are similar to those of a one-dimensional configuration, side-wall wetting acts as a stabilizing mechanism for long-wave perturbations, leading to a significant increase in the instability threshold Reynolds number. 

To examine the transition between these distinct roles of wetting across different confinement levels, figure \ref{phase} presents a phase diagram showing the wetting-induced shift of the stability threshold as a function of the confinement ratio $W/H$ and the wavenumber $k$. The color contours represent the relative change in the critical Reynolds number due to wetting:

\begin{equation}\label{wet_shift}
	\Delta_{\mbox{wet}} = \frac{Re_c^{\mbox{visc}} - Re_c^{\mbox{wet+visc}}}{Re_c^{\mbox{visc}}} \times 100,
\end{equation}
where $Re_c^{\mbox{visc}}$ denotes the critical Reynolds number in the non-wetting confined case, and $Re_c^{\mbox{wet+visc}}$ denotes the corresponding value when wetting is included. Wetting is destabilizing relative to the non-wetting confined flow for positive values of $\Delta_{\mbox{wet}}$, while negative values indicate stabilization. Moreover, the top horizontal axis shows the corresponding channel width to capillary length ratio, obtained as follows:

\begin{equation}\label{ratios}
	\frac{W}{l_c} = \frac{W}{H} \sqrt{\frac{Ca}{\tan{\beta}}},
\end{equation}
where $Ca$ and $\beta$ are fixed. This axis provides a measure of the change in the spatial extent over which side-wall wetting forces act as confinement ratio is varied. Moreover, the dotted line corresponds to a threshold for viscous confinement stabilization defined as:

\begin{equation}
	\Delta_{\mbox{visc}} = \frac{Re_c^{\mbox{visc}} - Re_c^{\mbox{1D}}}{Re_c^{\mbox{1D}}} \times 100 = 2\%,
\end{equation}
indicating the boundary at which viscous side-wall confinement effects become negligible to within $2\%$. To the right of this line, the non-wetting flow closely approaches the one-dimensional stability limit.

To the left of the dotted line, viscous boundary layers at the side walls play an important stabilizing role. In  this region, wetting acts to weaken this stabilization effect, leading to a relative destabilization at moderate wavenumbers. The magnitude of this destabilization decreases as confinement is relaxed. At the same time, the influence of wetting on long wave perturbations remains comparatively weak.

As the confinement ratio increases, the stabilizing-influence of the viscous boundary layer is diminished and the wetting-induced destabilization at moderate wavenumbers fades. At the same time, despite the fact that the local strength of the side-wall wetting decreases as $W/l_c$ increases, capillary effects are able to act over an increasingly wide spanwise region. Consequently, wetting  emerges as a dominant mechanism controlling long-wave instability, leading to the pronounced stabilization at small wavenumbers. Subsequently, this stabilization  weakens as the confinement ratio is increased further. For sufficiently large confinement ratio ($W/H \gtrsim 190$),  the flow approaches the one-dimensional limit and both viscous confinement and wetting effects are diminished, consistent with the absence of any side-wall influence in the unconfined channel limit. 

Overall, the phase diagram demonstrates that side-wall wetting plays two distinct and competing roles depending on the confinement level. In confined channels, wetting counteracts the stabilizing influence of the side-wall viscous boundary layer at moderate wavenumbers and therefore acts as a relatively destabilizing mechanism. As confinement is relaxed and viscous stabilization becomes negligible, wetting acts as a stabilizing mechanism for long wave perturbations and becomes the dominant factor controlling the instability threshold.

\subsection{Long-wave instability threshold and confinement scaling}

\begin{figure}
	\centering
	\includegraphics{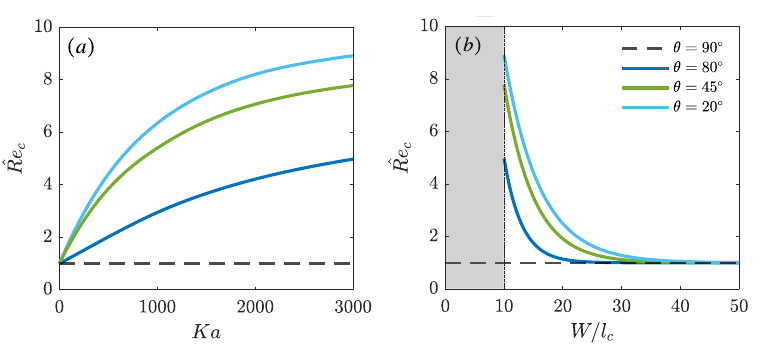}
	\caption{ Critical Reynolds number normalized by the classical one-dimensional critical Reynolds number   for different contact angles $\theta$ as a function of (a) the Kapitza number $Ka$ and (b) the channel-width-to-capillary-length ratio $W/l_c$.}
	\label{Re_c}
\end{figure}
The results of the previous sections indicate that side-wall wetting can significantly increase the long wave instability threshold even in channels that would traditionally be classified as non-confined based on the geometric confinement ratio $W/H$. To isolate the underlying mechanism, we investigate the stability of long-wave disturbances $(k=0.01)$ by computing the critical Reynolds number as a function of the Kapitza number $Ka$ and of the channel width to capillary length ratio $(W/l_c)$. In these computations, the geometric confinement ratio $W/H$ is held fixed, while the ratio $W/l_c$ varies with $Ka$ through its dependence on fluid properties. This parametrization  isolates the dependence of the instability threshold on $Ka$ at a prescribed geometric confinement level.

Figure \ref{Re_c}(a) shows the normalized critical Reynolds number $\hat{Re}_c = Re_c/Re_{c_{1D}}$ as a function of $Ka$ for different contact angles.  In the non-wetting limit, ($\theta= 90^{\circ}$), $\hat{Re}_c$ is essentially independent of $Ka$ and matches the classical one-dimensional prediction. This confirms that the delay in the instability threshold cannot be attributed to surface tension effects in the bulk alone but requires side-wall wetting effects, consistent with the physical mechanism proposed in the experiments.    In contrast, when wetting is present ($\theta  < 90^{\circ}$), $\hat{Re}_c$ increases strongly with $Ka$, where this effect intensifies as the contact angle decreases, reflecting a progressively  stronger stabilizing influence of wetting.  The observed dependence on $Ka$ is therefore a result of a coupling between capillarity and side-wall wetting, rather than from surface tension in the bulk.

Furthermore, figure  \ref{Re_c}(b) shows the variation of $\hat{Re}_c$ with $W/l_c$. For large $W/l_c$ the instability threshold approaches the one-dimensional value $Re_{c_{1D}}$, indicating that the influence of side walls becomes negligible as the capillary length becomes small compared to the channel width. As $W/l_c$ decreases, the instability threshold increases rapidly, with a stronger rise for smaller contact angles. The grey-shaded region indicates $W/l_c <10$, corresponding to a regime in which the capillary elevations from the two side walls overlap and the entire spanwise direction is reshaped by capillarity; this regime is outside the scope of this study. We therefore restrict our analysis to $W/l_c \ge 10$, where the central part of the film remains close to the one-dimensional Nusselt solution \citep{nusselt1916oberflachenkondensation}. The rapid rise in $\hat{Re}_c$ indicates the onset of a strong stabilization, rather than a true singular limit as $W/l_c \to 10$.

\subsection{Comparison with experimental measurements}
\begin{figure}
	\centering
	\includegraphics{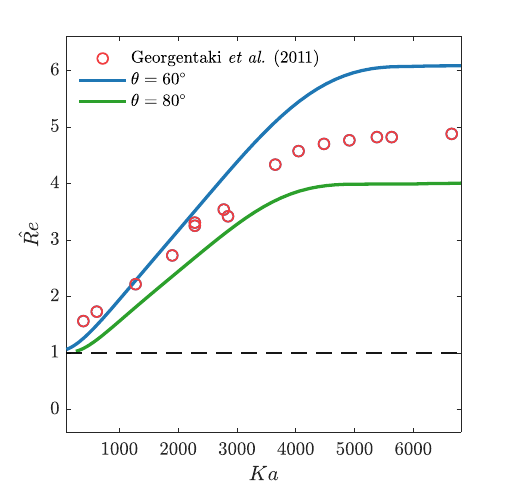}
	\caption{ Comparison between theoretical predictions and experimental measurements of the normalized critical Reynolds number as a function of the Kapitza number for the channel $W=100$ mm of \cite{georgantaki2011measurements}. }
	\label{exp_theor}
\end{figure}

In this section, we assess the quantitative consistency of our theoretical stability predictions with the experimental measurements of \cite{georgantaki2011measurements}. In those experiments, the influence of side-wall wetting on the long-wave instability was investigated by measuring the shift of  the instability threshold as the Kapitza number increases for a fixed channel width, using aqueous solutions of glycerol. It was shown that the critical Reynolds number increases significantly with $Ka$ and, at sufficiently large $Ka$, reaches a plateau whose value depends on the channel width. This plateau value was shown to be correlated with the channel width to capillary length ratio $W/ l_c$ through the relation $Re_c/Re_{c_{1D}} = 1 + 125/ (W/ l_c)$.  

In the experiments, $Ka$ is varied primarily through changes in viscosity. Over the range of mixtures considered, density and surface tension vary significantly less than viscosity, so the capillary length defined in \eqref{capillary_length}, and hence $W/l_c$ varies weakly for a fixed channel width. To approximate this experimental configuration, we fix $W/l_c$ at the value inferred from the reported high-$Ka$ plateau via the above correlation. Under this constraint, the film thickness $H$, and hence the ratio $W/H$ are not imposed independently; instead, it follows from the scaling relation in \eqref{ratios}, once $Ka$ and neutral $Re$ are determined.

The static contact angle of the aqueous solutions on Plexiglas (PMMA) sidewalls is not reported in \cite{georgantaki2011measurements}. We therefore compare the experimental measurements against an envelope of theoretical predictions obtained by varying the static contact angle over a representative partially wetting range.   Reported contact angles of water and glycerol on PMMA typically lie in the range $65^{\circ} - 75^{\circ}$ \citep{abdel2019surface}.  We therefore consider an interval of  $[60^{\circ} - 80^{\circ}]$ in order to account for uncertainties resulting from surface preparation and mixture compositions. 

Figure \ref{exp_theor} compares the theoretical predictions with the experimental measurements for the $W=100 \mbox{mm}$ channel, for which the plateau correlation yields $W/l_c \approx 30$. The calculations are performed at a wavenumber $k=0.01$, and inclination angle $\beta=10$. The resulting geometric confinement ratio satisfies $W/H>100$, consistent with the experimental regime.  The theoretical predictions reproduce the experimentally observed monotonic increase of the critical Reynolds number with $Ka$, as well as the saturation at larger values.  The experimental data fall within the theoretical envelope over the measured $Ka$ range.  Overall, the agreement provides quantitative support for the present theoretical framework and its interpretation of the wetting-induced mechanisms presented in the  previous sections.

\subsection{Perturbation mode shape}

\begin{figure}
	\centering
	\includegraphics{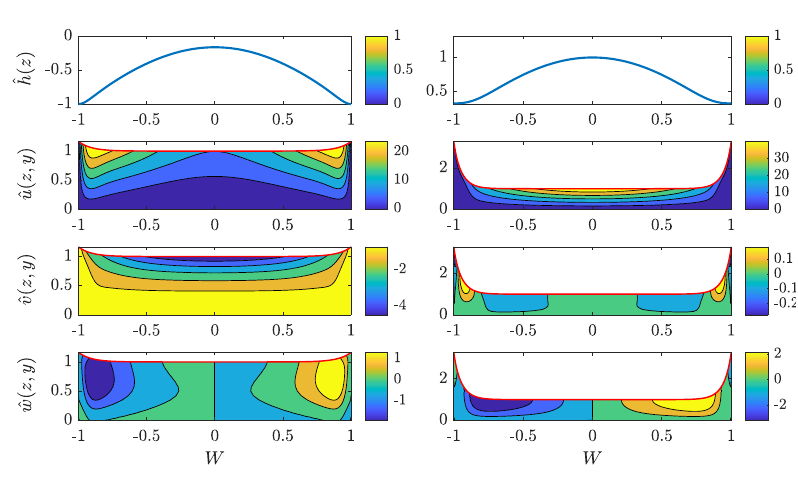}
	\caption{  Perturbation eigenmode structures.  Left:  relatively destabilizing case ($k=0.4$, $\theta = 80^{\circ}$, $Ca=0.2$). Right: stabilizing case ($k=0.01$, $\theta = 45^{\circ}$, $Ca = 0.02$ ). Panels show the real part of the phase-aligned eigenfields $\hat{h}(z),\hat{u},\hat{v},\hat{w}$ . For each mode (column), all fields are scaled by the same factor $\max|\hat{h}|$ so that the color scale is consistent.   }
	\label{modes}
\end{figure} 

To gain a deeper insight into how wetting weakens the confinement-induced stabilization in confined channels, or stabilizes the long-wave perturbations in weakly-confined channels, we examine the eigenmode amplitude fields for two representative cases. Figure \ref{modes} shows the interface perturbation amplitude $\hat{h}(z)$, and the velocity perturbation amplitudes $\hat{u},\hat{v},\hat{w}$ for a relatively destabilizing  short-wave mode in a confined channel (left, $W/H=20$), and a stabilizing effect on a long-wave perturbation in a weakly-confined channel (right, $W/H=100$). 

In the relatively destabilizing case in confined channels (left), the interface perturbation  $\hat{h}$ peaks at the center but decays towards the side walls with steep lateral gradients and without a notable flattening at the boundaries, indicating a weaker pinning effect caused by wetting. More importantly, both  $\hat{u}$ and  $\hat{w}$ develop intense structures within the meniscus region, indicating that the disturbance is strongly influenced by the near-wall region. 

In contrast, for the stabilizing case (right), the interface perturbation remains at a maximum at the channel center, but approaches smaller values near the side walls, consistent with a stronger wall-anchoring of the disturbance shape.  The velocity perturbations are predominantly organized away from the meniscus region. The streamwise component $\hat{u}$ is mainly concentrated in the core, $\hat{v}$ shows relatively small fields near the boundaries, and $\hat{w}$ forms a pair of counter-rotating cells that fill most of the span, with their center located outside the meniscus region. 

Overall, wall-anchoring of the interface together with span-filling secondary-flow structures correlates with a stabilizing influence of wetting, while the absence of interface anchoring and the presence of vortical structures in the meniscus region correlate with a relatively destabilizing effect.  Furthermore, these modal structures relate directly to the experimental observations of \cite{georgantaki2011measurements}, who inferred the existence of a recirculating motion in the streamwise-spanwise plane where liquid is transported  from the channel center towards the side walls along the wave crest, and returns from the side walls towards the center at a different streamwise phase.  The spanwise vortices associated with the eigenmodes in figure \ref{modes} represent the onset of this  secondary motion and the corresponding spanwise redistribution of liquid, driven by lateral variations of film thickness and interface curvature between the channel center and lateral menisci.

\section{Conclusions}\label{sec:conclusion}
In this work, we investigate the effect of side-wall wetting on the linear stability of gravity-driven liquid films confined in the spanwise direction. In such a configuration, the interplay between the classical hydrodynamic instability, spanwise confinement and capillarity associated with side-wall wetting has remained largely unexplored. We develop a biglobal stability framework based on the linearized Navier--stokes equations, regularized by a Navier slip condition at the side walls to resolve the moving contact line singularity. The base state solution is obtained by mapping the curved physical domain onto a rectangular computational grid, and the resulting meniscus shape and near-wall velocity overshoot are  verified against the  experiments of \citep{haas2011side}.

We consider two limiting regimes. In confined channels ($W/H=20$), spanwise confinement alone provides a significant stabilizing influence at moderate wavenumbers. When side-wall wetting is introduced, the computed \( Re\text{--}k \) stability contours reveal that wetting substantially reduces the confinement-induced stabilization effect, and in some cases, can nearly cancel it. Therefore, wetting plays a relatively destabilizing role with respect to the non-wetting confined case.   Analysis of the associated base state shows that this behavior is linked to a streamwise velocity overshoot region near the side walls, which  weakens  the effective viscous damping provided by the side-wall boundary layer.

In contrast,  in weakly-confined channels ($W/H=100$), spanwise confinement plays no significant role in the absence of wetting, and the classical one-dimensional stability behavior is recovered. When wetting is introduced,  \( Re\text{--}k \)  stability contours show little change at moderate wavenumbers, but have a pronounced effect on long-wave perturbations $k \rightarrow 0$. The instability threshold is shifted to higher Reynolds numbers, with the onset increasing to larger $Re$ as wetting becomes stronger.  

A phase diagram quantifying the wetting-induced modification of the instability neutral curves shows that the crossover from the relative destabilization to long-wave stabilization occurs smoothly over a finite range of $W/H$, rather than a sudden regime switch.  Moreover, we provide a quantitative comparison with the  experimental measurements of \cite{georgantaki2011measurements} based on the critical Reynolds number of long-wave perturbations. The experimental data fall within a physically consistent envelope of theoretical predictions and show similar trends and comparable magnitudes within the accessible parameter range. 

Furthermore, the mechanisms behind the distinct relative destabilization and stabilization effects are clarified by examining the perturbation eigenmode structures. In confined channels, wetting introduces vortical structures that are localized in the meniscus region, associated with  streamwise velocity overshoot, which weaken the confinement-induced stabilization. In contrast, in weakly-confined channels, the velocity perturbations remain concentrated in the center of the channel, while the interface perturbation exhibits strong anchoring at the side walls, consistent with a net stabilization driven by interface "tensioning" due to wetting.

\section{Future work}
A nonlinear stability analysis constitutes a natural continuation of this study, as it would capture the finite-amplitude dynamics of the moving contact line, and the nonlinear interaction between different unstable modes.  In particular, nonlinear contact line dynamics may introduce additional dissipation or coupling effects that are not captured within the present linear framework. Such an analysis would provide a more comprehensive understanding of the influence of wetting beyond the onset of the instability.

Another promising direction concerns the interaction of wetting with thermal instabilities.  Thermocapillary \citep{mohamed2024stability} and vapor instabilities \citep{mohamed2020linear}  exhibit  inherently three-dimensional structures with significant spanwise modulation. It would therefore be valuable to investigate how these modes interact with side-wall wetting and geometric confinement within a similar biglobal framework. 

The present study is restricted to wetting conditions with contact angles $\theta \le 90^\circ$, for which capillary elevation near the side walls leads to local thickening of the film. A further extension concerns conditions with $\theta > 90^\circ$ where the wetting behaviour is reversed and capillary depression leads to a local decrease in the film thickness near the side walls. This modification of the base state structure is expected to affect the stability characteristics. A systematic investigation of this regime remains to be carried out.

\appendix
\section{Coordinate transformation}  \label{Appendix_B}
The physical domain in this work is not rectangular due to the presence of the curved meniscus at the side walls.  Therefore, the physical domain with coordinates bounded as ($y \in [0, \bar{h}(z)]$, $z \in [-W,W] $) is mapped into the Chebyshev domain  ($y_c \in [-1,1]$ and $z_c \in [-1,1]$) using the following transformations: 

\begin{subequations} 
	\begin{align}
		&  	z = W z_c,  \\
		&  	y = \frac{\bar{h}(z)}{2} (y_c + 1),
	\end{align}
\end{subequations}
where  $\bar{h}(z)$ is the base state interface shape defining the top boundary of the domain. For an arbitrary function $\phi(z,y)$, the partial derivative in the physical domain can be related to the partial derivatives in the computational domain using the following transformations \citep{john1995computational}: 

\begin{subequations} 
	\begin{align}
		&    \frac{\partial \phi}{\partial y} = \frac{\partial y_c}{ \partial y} \frac{\partial \phi}{\partial y_c},  \\
		&    \frac{\partial  \phi}{\partial z} = \frac{\partial z_c}{\partial z} \frac{\partial \phi}{\partial z_c} +  \frac{\partial y_c}{\partial z} \frac{\partial \phi}{\partial y_c}, \\
		&    \frac{\partial^2 \phi}{\partial y^2} = \Big(\frac{\partial y_c}{\partial y}\Big)^2 \frac{\partial^2 \phi}{\partial y_c^2},  \\
		& \frac{\partial^2 \phi}{\partial z^2} = 	\Big(\frac{\partial z_c}{\partial z}\Big)^2 \frac{\partial^2 \phi}{\partial z_c^2} + 2\,\frac{\partial z_c}{\partial z}\frac{\partial y_c}{\partial z} \frac{\partial^2 \phi}{\partial z_c \partial y_c} + \Big(\frac{\partial y_c}{\partial z}\Big)^2 \frac{\partial^2 \phi}{\partial y_c^2} + 	\frac{\partial^2 y_c}{\partial z^2}\frac{\partial \phi}{\partial y_c} +  \frac{\partial^2 z_c}{\partial z^2}\frac{\partial \phi}{\partial z_c}  \\
		&    \frac{\partial ^2 \phi}{\partial z \partial y} =  \frac{\partial^2 y_c}{\partial z \partial y} \frac{\partial \phi}{\partial y_c} + \frac{\partial y_c}{\partial z} \frac{\partial y_c}{\partial y} \frac{\partial^2 \phi}{\partial y_c^2} + \frac{\partial z_c}{\partial z} \frac{\partial y_c}{\partial y}  \frac{\partial^2 \phi}{\partial z_c \partial y_c}. 
	\end{align}
\end{subequations}
where the following expressions define the derivatives of the computational coordinates with respect to the physical coordinates:

\begin{subequations} 
	\begin{align}
		&     \frac{\partial z_c}{\partial z} = \frac{1}{W}, \qquad  \qquad \qquad  \frac{\partial y_c}{\partial y} = \frac{2}{\bar{h}(z)}, \qquad  \qquad \qquad \frac{\partial y_c}{\partial z} = \frac{-2 y}{\bar{h}^2} \frac{d\bar{h}}{dz},\\
		& \frac{\partial^2 y_c}{\partial z^2} = \frac{4 y}{\bar{h}^3} \Big( \frac{d\bar{h}}{dz}   \Big)^2 -  \frac{2y}{\bar{h}^2}  \frac{d^2 \bar{h}}{dz^2}, \qquad \qquad  \qquad \qquad \qquad \frac{\partial^2 y_c}{\partial z \partial y} =   \frac{-2}{\bar{h}^2}  \frac{d \bar{h}}{dz}.  
	\end{align}
\end{subequations}

\section{ Grid convergence and slip-length sensitivity} \label{slip_length}
The relation between the slip length $l_s$ and the spatial resolution is a key numerical consideration. When the slip length is sufficiently larger than the smallest grid spacing, its effect on the solution is properly resolved, whereas when it is  smaller  than the grid spacing, the solution becomes numerically indistinguishable from the no-slip limit.  

\begin{figure}
	\centering
	\includegraphics{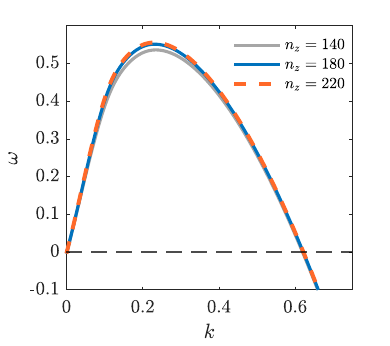}
	\caption{Grid convergence of the temporal growth rate for a confined channel with strong wetting ($W/H=20$, $W/l_c=10$, $\theta=15^{\circ}$, $Re = 50$, and $\beta=10^{\circ}$). Results are shown for three spanwise resolutions, $n_z = 140, 180, \mbox{and } 220$, corresponding to minimum grid spacing $\Delta z_{min}  = 2.5 \times 10^{-3}, 1.5 \times 10^{-3}, \mbox{and } 1 \times 10^{-3}$, respectively. }
	\label{grid_conv}
\end{figure}

We first assess the grid convergence for a fixed non-dimensional slip length  $l_s = 10^{-2}$, which is the value used throughout the manuscript. The convergence test is performed for a representative and numerically demanding configuration in which confinement and wetting effects are particularly strong, namely $W/H=20$, $W/l_c=10$, and $\theta=15^{\circ}$. Figure \ref{grid_conv} compares the growth rate curves obtained using several grid resolutions in the spanwise direction. The results show that further grid refinement leads to negligible changes in the growth rate curves and dominant eigenvalues, demonstrating convergence with respect to spatial resolution. The finest grid shown ($n_z = 220$) yields a minimum grid spacing near the side walls of $\Delta z_{min} = 1 \times 10^{-3}$, which is smaller than the chosen slip length, ensuring that the slip boundary condition is well resolved. Unless otherwise stated, this resolution is chosen in the remainder of the study. 

\begin{figure}
	\centering
	\includegraphics{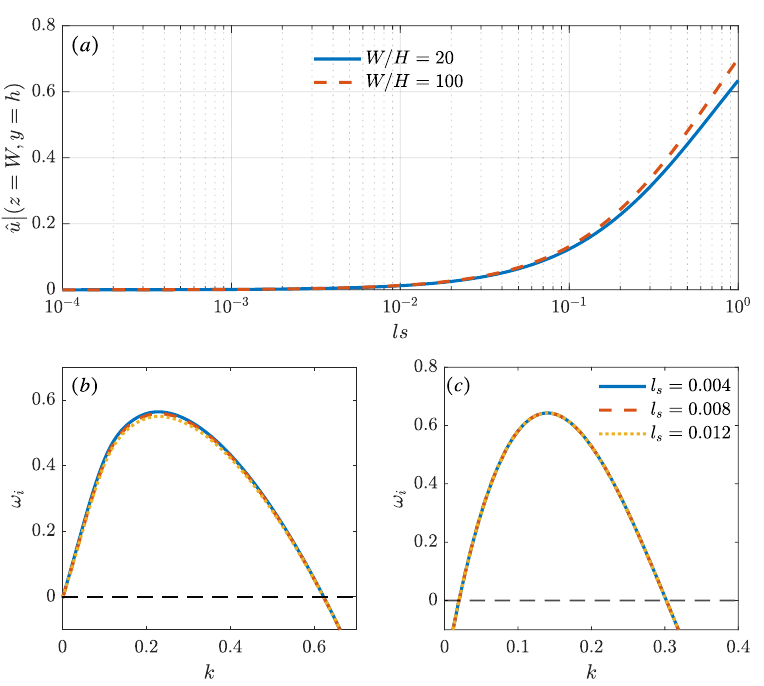}
	\caption{Effect of slip length on the base state and linear stability.  (a) Normalized base-state interface velocity at the side wall as a function of $l_s$.  (b,c) Temporal growth-rate curves for (b) a confined channel ($W/H=20$) and (c) a weakly confined channel ($W/H=100$) for several slip lengths.  Unless otherwise stated, $Re=100$, $W/l_c=10$, $\theta=15^\circ$, and $\beta=10^\circ$.}
	\label{ls}
\end{figure}

Next, we examine the sensitivity of the results with respect to the slip length on the converged grid. Figure \ref{ls}(a) shows the base state interface velocity  at the side walls, normalized by the velocity at the channel center, as a function of the slip length for both confined and weakly confined channel configurations, with the grid spacing  fixed to $\Delta z_{\min} \approx 10^{-3}$. For ${l}_s \lesssim 10^{-3}$, the wall velocity collapses onto the  no-slip limit, reflecting the fact that such small slip lengths are under-resolved.  As the slip length increases and becomes several times larger than the grid spacing, its influence becomes apparent as the interface velocity increases  approximately logarithmically with  $l_s$. This behavior is consistent with  the classical contact line theory, in which the slip length enters macroscopic predictions  through logarithmic dependence \citep{voinov1976hydrodynamics,cox1986dynamics}. Importantly, the base state response to slip length is found to be very similar in confined and weakly confined channel configurations, indicating that the slip length condition has a local effect, rather than a global effect on the whole channel width.

Furthermore, figures  \ref{ls}(b,c) show the corresponding linear stability results for several slip length values. Panel (b) represents the confined channel configuration ($W/H = 20$), where the side wall boundary layers play an important role. In this case, changing the slip length, while keeping it sufficiently resolved by the grid, results in a modest shift in the growth rate curves, most noticeably at moderate wavenumbers, reflecting a weakening of the confinement-induced viscous damping. The maximum variation in the growth rate   between the smallest and largest slip length remains below $2\%$.  Additionally, panel (c) shows the same analysis for a weakly-confined channel configuration ($W/H = 100$). In this regime, the growth rates for different slip lengths are indistinguishable on the scale of the figure over the entire wavenumber range, including the small-wavenumber stabilization band discussed previously in figure \ref{omega_k_for_G}. This confirms that the stabilization mechanism at small wavenumbers in weakly-confined channels is not affected by near-wall slip regularization.  

Overall, the aforementioned results demonstrate that linear stability characteristics depend weakly on the slip length once it is sufficiently resolved on the grid. The modest sensitivity observed in the confined configuration does not change the qualitative stability behavior or the main findings of this study. Consequently, unless otherwise stated, all the results reported in this study are obtained using  a minimum grid spacing of $\Delta z_{\min} \approx 10^{-3}$ and non-dimensional slip length of $l_s=0.01$. 

\noindent\textbf{Declaration of interests.} The authors report no conflict of interest.

\bibliographystyle{jfm}
\bibliography{jfm}

\end{document}